%% file: czechmishrainggs.tex
\documentclass[journal]{IEEEtran}
\usepackage{graphicx}
\usepackage{epsfig}
\usepackage{epstopdf}
\usepackage{array}
\usepackage{amsmath}
\usepackage{amssymb}
\usepackage{etoolbox}
\usepackage{tabularx}
\usepackage{booktabs}
\usepackage{balance}
\usepackage{caption} 
\begin{document}
\bstctlcite{IEEEexample:BSTcontrol}

\title{A CNN and LSTM-Based Approach to Classifying Transient Radio Frequency Interference}

\author{Daniel~Czech\textsuperscript{1},
        Amit~Mishra
        and~Michael~Inggs% 

\thanks{D. Czech, A. Mishra and M. Inggs are with the Department of Electrical Engineering, University of Cape Town.}% <-this % stops a space
\thanks{\textsuperscript{1}Corresponding author: daniel.josef.czech@gmail.com}}

\maketitle

\begin{abstract}

Transient radio frequency interference (RFI) is detrimental to radio astronomy. It is particularly difficult to identify the sources of transient RFI, which is broadband and intermittent. Such RFI is often generated by devices like mechanical relays, fluorescent lighting or AC machines, which may be present in the surrounding infrastructure of a radio telescope array. 
One mitigating approach is to deploy independent RFI monitoring stations at radio telescope arrays. Once the sources of RFI signals are identified, they may be removed or replaced where possible. 
For the first time in the open literature, we demonstrate an approach to classifying the sources of transient RFI (in time domain data) that makes use of deep learning techniques including CNNs and LSTMs. 
Applied to a previously obtained dataset of experimentally recorded transient RFI signals, our proposed approach offers good results.
It shows potential for development into a tool for identifying the sources of transient RFI signals recorded by independent RFI monitoring stations.

\end{abstract}


\IEEEpeerreviewmaketitle

\section{Introduction}

\IEEEPARstart{R}{adio} astronomy continues to face the problem of radio frequency interference (RFI). As instruments become more sensitive, so the impact of existing RFI sources becomes more significant. New technologies that make use of the RF spectrum become more widely adopted over time. To counter the growing problem of RFI, a variety of approaches have been developed and refined. 

Most commonly, RFI is detected in data directly from radio telescopes. Such approaches typically distinguish only between RFI and astronomical signals, making no attempt to determine the identity of the sources of RFI signals. A wide variety of algorithms have been developed, mostly for application with 2D time-frequency data \cite{mitigation2, mitigation1}.

An additional approach, one being employed at the Square Kilometer Array (SKA) site in South Africa, is to develop independent RFI monitoring stations \cite{monitor2, monitor1, monitor3}. These stations will continuously monitor almost the full bandwidth of the radio telescope, simultaneously and in all directions. They also have the ability to capture time-domain transient RFI signals. Such monitoring stations make it easy to detect nearby sources of RFI, so that they may be removed or replaced.
 
In the case of intentional transmissions (for example, telecommunications) it is usually easy to identify their sources since they adhere to government-allocated frequency bands. Transient RFI signals are much harder to identify, however, since they are intermittent and broadband. Typically, they are produced as a byproduct of the normal operation of devices such as mechanical relays, fluorescent lights, AC machines etc. 

There are few prior attempts to identify the sources of transient RFI in a radio-astronomy context. Unsupervised clustering via the k-means algorithm was applied to transient RFI at the Parkes radio telescope \cite{doran2013}, but individual sources were not classified. In other work \cite{wolfaardt2016} a variety of supervised learning techniques were employed to classify sources of RFI in labelled data recorded at the MeerKAT construction site in South Africa. Gaussian mixture model and k-nearest neighbours classifiers were applied to the data. High classification accuracy was achieved, however the number of samples per class was very small (in some cases less than 10). 

In our own prior work, we looked at classifying RFI events using nonlinear principal components analysis techniques \cite{czech2017a} as well as a dictionary-based approach in conjunction with hidden Markov models \cite{czech2017b}. Signals were recorded from a number of common sources of transient RFI under controlled conditions, using a custom capturing system very similar to those which are installed in RFI monitoring stations at the MeerKAT/SKA site in South Africa. 

Attempts have been made in other fields to classify similar types of transient RF signal. In one such approach, basic neural networks were used to classify the makes of different vehicles based on their transient RF emissions \cite{d2006}.

In this paper, we propose a novel approach to classifying the sources of transient RFI. Recurrent neural networks, in particular, long short-term memory (LSTM) networks \cite{h1997} have proven highly effective at modeling time-dependent signals in a variety of applications, for example phoneme classification in automated speech recognition \cite{graves2005} and acoustic modeling \cite{peddinti2017}.

While they are known best for their use in visual processing, convolutional neural networks (CNNs) have also shown success in dealing with temporal sequence data, for example human speech \cite{sainath2013} and wireless interference identification for coexistence management \cite{schmidt2017}. A CNN-based approach has been used to identify sources of interference in WiFi signals \cite{longi2017}, although most of the sources dealt with were intentional continuous transmitters. Recordings were limited to the WiFi band and were recorded as time-frequency data. In a radio astronomy context, CNNs have been used to flag RFI in data from radio telescope arrays in recent work \cite{akeret2017}, but not to classify the flagged RFI by source. In addition, they were applied to data represented in the 2D time-frequency domain. 

The main novelties of our work consist of the following: As far as we are aware, this is the first time that either CNNs or LSTMs have been employed to identify the sources of time-domain transient RFI. In addition, we believe our approach combining 1D-CNNs and bidirectional LSTMs has not been attempted elsewhere for the classification of RFI signals (of any type) by their sources. This work constitutes one of the relatively few attempts available in the open literature thus far to identify the sources of transient RFI signals.     

The rest of this paper is organised as follows: The experimental data and associated preprocessing steps are described in Section~\ref{data}. The models and and their application to the data are discussed in Section~\ref{model}. In Section~\ref{results}, results are presented. Finally, conclusions are drawn in Section~\ref{conclusion}.

\section{Data and Preprocessing}
\label{data}
\begin{figure*}[t]
	\centering
	\includegraphics[width=15cm]{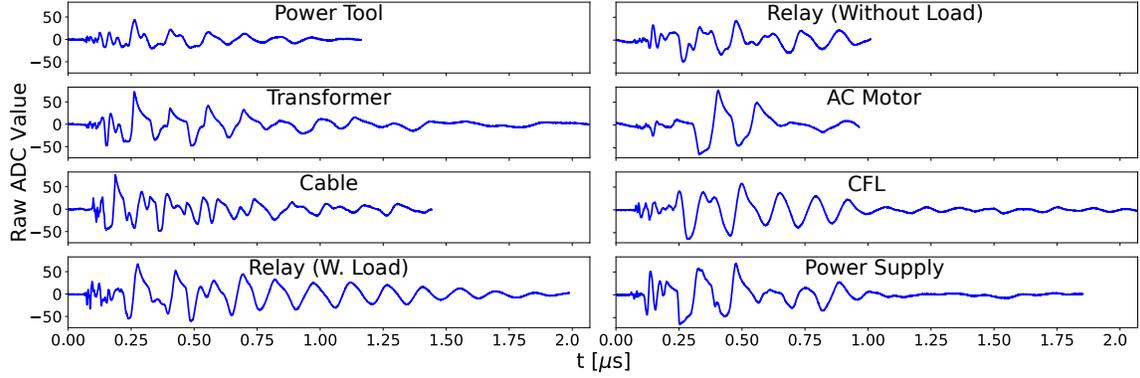}
	\caption{Examples of individual transients from each class as extracted by the automated algorithm given in \cite{czech2017b}. The lengths of the transients differ to an extent.}
	\label{example_signals}
\end{figure*}

The data used in this analysis are derived from our prior work \cite{czech2017a}. In the original dataset, full RFI recordings consist of a sequences of individual transients. In other prior work \cite{czech2017b} we presented an algorithm for extracting these individual transients from full RFI recordings. In this paper, we use these individual transients, extracted using this algorithm. This dataset consists of 63130 individual transients from 8 different sources. An example of a transient RFI signal from each class is given in Fig.~\ref{example_signals}. Transients are aligned by their largest peaks, and padded with zeros where necessary (since their lengths vary).

This data format, time-domain captures of short transient RFI signals, is one of which independent RFI monitoring stations at the SKA site in South Africa will record. The ability to identify the sources of transient RFI, as recorded by such monitoring stations, would be highly valuable.

\subsection{Preprocessing}

Prior to classification, limited preprocessing steps are carried out. Each transient is limited in length to 5000 raw samples for two reasons: One, the majority of transients are shorter than 5000 samples, and two, unnecessary computational overhead is avoided. The amplitude range of each individual transient (for all train, test and validation sets) is scaled to range between -1 and 1: for each transient $t$, $t_{scaled} = 2\frac{t - min(t)}{max(t)-min(t)}-1$. This scaling is applied because ideally, an RFI classification system would be capable of handling variations in amplitude. For example, in the field, the amplitudes of RFI signals will vary according to the distance from their sources. Transients are also standardised across each feature (time step). For each feature vector $x_j$ containing one value from each sample in a training set, $x_{j(scaled)} = \frac{x_j - \mu_j}{\sigma_j}$. This ensures that each feature is no more influential than the next. The standardisation parameters are determined from the training data alone; these predetermined parameters are used when standardising validation and test data. 

The data are split into training, validation and testing sets. The training set consists of 60\% of the available data, while the others account for 20\% each. The data is stratified by class (each set contains an equal proportion of samples from each class) and shuffled (so the order of the samples in each set is random). The validation set is used for hyperparameter tuning, while the test set is kept separate until final evaluation, where the training set consists of both the original training and validation sets combined. 

\subsection{Class Imbalance}
\label{imbalance}

Certain RFI sources (such as the mechanical relay) produced many more transients in a single event sequence than others. As a result, the number of individual transients is significantly imbalanced by class. The number of transients (equivalently, samples) per class is given in Table~\ref{devices_table}. Due to the class imbalance, we perform two separate analyses. In the first approach, we balance the classes by limiting the number of samples per class to the number of samples in the smallest class. For the larger classes, a subset of their samples is drawn at random. In the second approach, rather discarding data, samples are weighted by class in the cost function. Samples from rarer classes are weighted higher than samples from common classes, ensuring that each class has an equivalent influence on the model during training.

\begin{table}[h]
	\renewcommand{\arraystretch}{1.4}
	\centering
	\caption[]{RFI sources}
	\vspace{4mm}
	\begin{tabular}{>{\centering\arraybackslash}m{6mm}m{45mm} m{18mm}}
		\hline
		\textbf{Class} & \textbf{Description} & \textbf{No. Samples}\\
		\hline
		1 & Compact fluorescent lamp  & 662\\
		2 & Power tool & 543 \\
		3 & Step-down transformer & 5523  \\
		4 & Cable & 264 \\
		5 & Mechanical relay (700W resistive load) & 16006 \\
		6 & Mechanical relay (without load) & 35932 \\
		7 & AC motor (approximately 1 kW) & 3675 \\
		8 & Small switching power supply & 525 \\
		\hline
	\end{tabular}
	\label{devices_table}
\end{table}

\section{Model Architecture}
\label{model}
\begin{figure}[t]
	\centering
	\includegraphics[width=4.6cm]{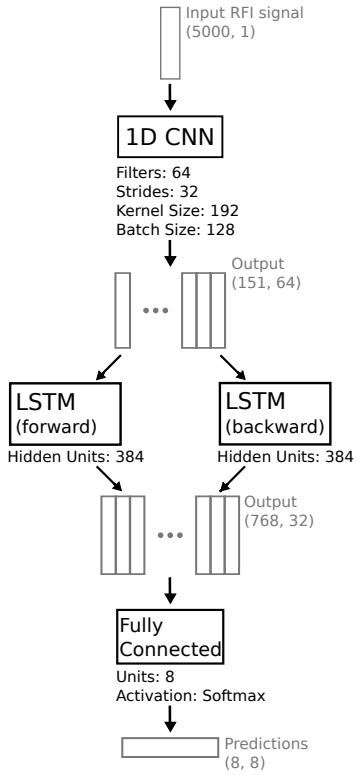}
	\caption{The architecture of the chosen model. In the bidirectional LSTM layer, the outputs of each LSTM are concatenated. The particular values given here apply for the balanced subset of the full dataset. Parameters were changed in some cases when the full unbalanced dataset was used: The CNN's pre-training batch size was increased to 256, while the kernel size was reduced to 160 time-steps.}
	\label{architecture}
\end{figure}

The architecture we selected is relatively uncomplicated - consisting of a 1D convolutional layer, followed by a bidirectional LSTM layer and finally a fully-connected layer, presenting the output in a 1-hot configuration. Fig.~\ref{architecture} illustrates the chosen model. The 1-D CNN layer serves both to identify salient features in the transient signals, and to reduce the length of the time-dependent input sequence to the LSTM layer. We chose to use LSTMs since they have proven highly effective at modeling temporal sequences in a wide variety of fields. In particular, bidirectional LSTMs have in some cases proven superior to other architectures in applications such as automated speech recognition, for example \cite{graves2005, zeyer2017}.

The hyperparameters for the CNN and LSTM layers were selected by training different configurations on the training set, and evaluating them on the validation set. Hyperopt \cite{hyperopt}, a Python library, was used to automate the hyperparameter selection process. Model training was carried out using the Python library Keras \cite{keras} with Tensorflow \cite{tensorflow}. Computations were performed using an Amazon p2.xlarge instance (2.7 GHz Broadwell CPU; 61 GiB RAM; 12 GiB NVIDIA Kepler K80 GPU).  

\begin{figure}[t]
	\centering
	\includegraphics[width=6.5cm]{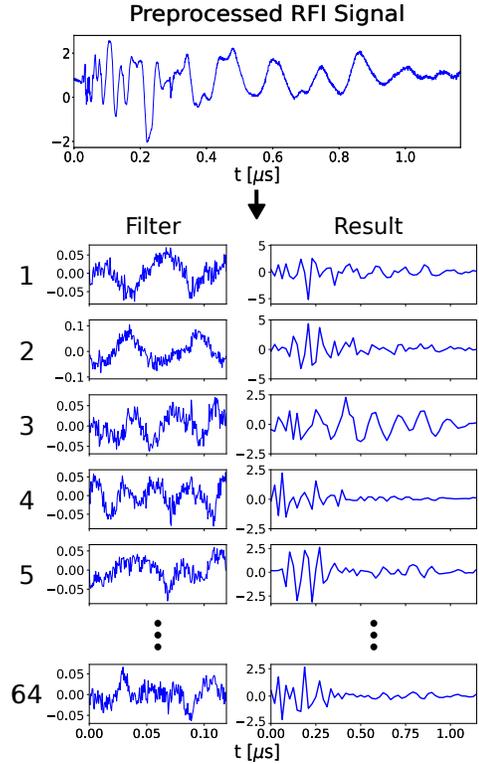}
	\caption{Several of the CNN's 64 filters and examples of their outputs when applied to a single preprocessed transient signal.}
	\label{filters}
\end{figure}

Model training was accomplished in two stages. Firstly, the CNN was pre-trained by replacing the LSTM layer with a temporary fully-connected classification layer. Next, the weights and filters obtained were kept fixed, and the temporary fully-connected layer replaced with the LSTM layer and a new, final fully-connected classification layer. Fig.~\ref{filters} shows 6 of the CNN's 64 filters. Some of the filters, such as those labelled 1, 2 and 4 suggest sinusoids of differing frequency and phase, while others such as 5 and 64 approximate other structural features. 

As discussed in Section~\ref{imbalance}, two approaches were taken when dealing with the dataset's class imbalance. In the first approach, each class was cut down to the same size, selecting (at random) an equal number of samples for each. This reduced the total number of samples considerably. In the second approach, the full dataset was used, balancing classes by increasing the weighting of samples from rarer classes accordingly in the cost function. If $L$ is the vector containing the number of samples $L_i$ in each class $i$ then the vector of class weights $C$ is calculated as follows: 

$$C = \frac{max(L)}{L}$$
\noindent
To account for these weights using categorical cross-entropy, the cost function is altered as follows: For a batch size $N$ of 1-hot output vectors of length $M$:

$$\mathcal{L}(y, \hat{y}) = -\frac{1}{N} \sum_{i=1}^{N}\sum_{j=1}^{M} y_{ij} \log(\hat{y}_{ij}) C_{j}$$

\noindent
and since we are using the softmax function, the partial derivative with respect to the output of the final layer is simply:

$$\frac{\partial \mathcal{L}}{\partial o_{i}} = C\odot(\hat{y}_i - y_i)$$
\noindent
where $\odot$ indicates element-wise multiplication and $o_i$ is the activation output. The gradients of the rarer classes are promoted relative to those of the more common classes due to the class weighting. 

\section{Results}
\label{results}

Results are presented for both approaches to the problem of class imbalance in the dataset. For evaluation, each model was trained with both the training and validation sets together (amounting to 80\% of the data) and tested on the as-yet unseen test set. In the first approach, the number of samples per class is limited to the number of samples in the smallest class. A confusion matrix is given in Table~\ref{confusion1} and other accuracy metrics in Table~\ref{metrics}. Precision and recall are calculated for each class and the mean of each metric reported. For $M$ classes, precision $ = \frac{1}{M}\sum_{i = 1}^{M}\frac{\text{tp}_i}{\text{tp}_i + \text{fp}_i}$ and recall $ = \frac{1}{M}\sum_{i = 1}^{M}\frac{\text{tp}_i}{\text{tp}_i + \text{fn}_i}$ where $\text{tp}_i = $ true positives, $\text{fp}_i = $ false positives and $\text{fn}_i$ = false negatives for class $i$.

In the second approach, no samples are discarded. Rather, samples are weighted in the training cost function according to the rarity of their class. A second confusion matrix is provided in Table~\ref{confusion2} and accuracy metrics given in Table~\ref{metrics}. Despite class imbalance, even the smallest classes are well classified. For example, the smallest class is correctly classified 96.15\% of the time. The single class with the worst classification accuracy was still classified correctly 77.14\% of the time.   

\begin{table}[h]
	\renewcommand{\arraystretch}{1.4}
	\centering
	\caption[]{Evaluation of Results}
	\vspace{4mm}
	\begin{tabular}{m{11mm}m{22mm} m{40mm}}
		\hline
		\textbf{Metric} &\textbf{Approach 1}& \textbf{Approach 2}\\
		&(Limited class size)& (Classes weighted in cost-function)\\
		\hline
		Accuracy & 0.8413  & 0.9636\\
		Precision & 0.8475 & 0.8467 \\
		Recall & 0.8413 & 0.9138  \\
		\hline
	\end{tabular}
	\label{metrics}
\end{table}

\begin{table}
	\renewcommand{\arraystretch}{1.9}
	\caption[]{Confusion matrix for the unseen test-set when classes are balanced by discarding data.}
	\vspace{3mm}
	\begin{tabular}{ r|p{1.85mm}|p{1.85mm}|p{1.85mm}|p{1.85mm}|p{1.85mm}|p{1.85mm}|p{1.85mm}|p{1.85mm}| }
		\multicolumn{1}{r}{}
		&  \multicolumn{1}{c} {\rotatebox[origin=l]{90}{CFL}}
		&  \multicolumn{1}{c}{\rotatebox[origin=l]{90}{power tool}}
		&  \multicolumn{1}{c}{\rotatebox[origin=l]{90}{transformer}}
		&  \multicolumn{1}{c}{\rotatebox[origin=l]{90}{cable}}
		& \multicolumn{1}{c}{\rotatebox[origin=l]{90}{relay (load)}} 
		&  \multicolumn{1}{c}{\rotatebox[origin=l]{90}{relay}}
		&  \multicolumn{1}{c}{\rotatebox[origin=l]{90}{AC motor}}
		&  \multicolumn{1}{c}{\rotatebox[origin=l]{90}{PSU}}\\
		\cline{2-9}
		{\small CFL} & \textbf{\scriptsize 44} &\scriptsize 0 &\scriptsize 2 &\scriptsize 0 &\scriptsize 2&\scriptsize 0 &\scriptsize 2 &\scriptsize 2\\
		\cline{2-9}
		\small power tool &\scriptsize 2 & \textbf{\scriptsize38} &\scriptsize 1  &\scriptsize 1 &\scriptsize 0 &\scriptsize 1 &\scriptsize 0 &\scriptsize 9 \\
		\cline{2-9}
		\small transformer &\scriptsize 3&\scriptsize 0 &\textbf{\scriptsize36} &\scriptsize 1 &\scriptsize 0 &\scriptsize 4 &\scriptsize 1 &\scriptsize 7\\
		\cline{2-9}
		\small cable &\scriptsize 2 &\scriptsize 0 &\scriptsize 0& \textbf{\scriptsize49} &\scriptsize 0 &\scriptsize 1 &\scriptsize 0 &\scriptsize 0\\
		\cline{2-9}
		\small relay (load) &\scriptsize4 &\scriptsize 0 &\scriptsize 0 &\scriptsize 1 & \textbf{\scriptsize45}&\scriptsize 1 &\scriptsize 1 & \scriptsize0\\
		\cline{2-9}
		\small relay &\scriptsize 1 &\scriptsize 0 &\scriptsize 4 &\scriptsize 0 &\scriptsize 0&\textbf{\scriptsize47} &\scriptsize 0 &\scriptsize 0 \\
		\cline{2-9}
		\small AC motor &\scriptsize1 & \scriptsize0 &\scriptsize0 &\scriptsize 1 &\scriptsize 1 &\scriptsize 1 &\textbf{\scriptsize48}  &\scriptsize 0\\
		\cline{2-9}
		\small PSU &\scriptsize 1 &\scriptsize 3 &\scriptsize 3 &\scriptsize 1 &\scriptsize 0 &\scriptsize 1 & \scriptsize0 &\textbf{\scriptsize43} \\
		\cline{2-9}
	\end{tabular}
	\label{confusion1}
\end{table}

\begin{table}
	\renewcommand{\arraystretch}{1.9}
	\caption[]{Confusion matrix for the unseen test-set when classes weighted in the loss function.}
	\vspace{3mm}
	\begin{tabular}{r|@{}p{2.5mm}|@{}p{2.5mm}|@{}p{2.5mm}|@{}p{2.5mm}|@{}p{2.5mm}|@{}p{2.5mm}|@{}p{2.5mm}|@{}p{2.5mm}| }
		\multicolumn{1}{r}{}
		&  \multicolumn{1}{c} {\rotatebox[origin=l]{90}{CFL}}
		&  \multicolumn{1}{c}{\rotatebox[origin=l]{90}{power tool}}
		&  \multicolumn{1}{c}{\rotatebox[origin=l]{90}{transformer}}
		&  \multicolumn{1}{c}{\rotatebox[origin=l]{90}{cable}}
		& \multicolumn{1}{c}{\rotatebox[origin=l]{90}{relay (load)}} 
		&  \multicolumn{1}{c}{\rotatebox[origin=l]{90}{relay}}
		&  \multicolumn{1}{c}{\rotatebox[origin=l]{90}{AC motor}}
		&  \multicolumn{1}{c}{\rotatebox[origin=l]{90}{PSU}}\\	
		\cline{2-9}
		{\small CFL} & \textbf{\scriptsize~~~120} &\scriptsize~~~1 &\scriptsize~~~7 &\scriptsize~~~0 &\scriptsize~~~2&\scriptsize~~~0 &\scriptsize~~~0 &\scriptsize~~~2\\
		\cline{2-9}
		\small power tool & \scriptsize~~~1 & \textbf{\scriptsize~~~88} &\scriptsize~~~6  &\scriptsize~~~2 &\scriptsize~~~2 &\scriptsize~~~3 &\scriptsize~~~0 &\scriptsize~~~6 \\
		\cline{2-9}
		\small transformer &\scriptsize~~~9&\scriptsize~~~4 &\textbf{\scriptsize~1035} &\scriptsize~~~0 &\scriptsize~~~1 &\scriptsize~~~19 &\scriptsize~~~5 &\scriptsize~~~31\\
		\cline{2-9}
		\small cable &\scriptsize~~~0 &\scriptsize~~~1 &\scriptsize~~~1& \textbf{\scriptsize~~~50} &\scriptsize~~~0 &\scriptsize~~~0 &\scriptsize~~~0 &\scriptsize~~~0\\
		\cline{2-9}
		\small relay (load) &\scriptsize~~~43 &\scriptsize~~~4 &\scriptsize~~~9 &\scriptsize~~~2 & \textbf{\scriptsize~3117}&\scriptsize~~~13 &\scriptsize~~~13 &\scriptsize~~~0\\
		\cline{2-9}
		\small relay &\scriptsize~~~2 &\scriptsize~~~9 &\scriptsize~~166 &\scriptsize~~~0 &\scriptsize~~~18& \textbf{\scriptsize~6957} &\scriptsize~~~22 &\scriptsize~~~12 \\
		\cline{2-9}
		\small AC motor &\scriptsize~~~1 &\scriptsize~~~0 &\scriptsize~~~11 &\scriptsize~~~0 &\scriptsize~~~4 &\scriptsize~~~3 &\textbf{\scriptsize~~716}  &\scriptsize~~~0\\
		\cline{2-9}
		\small PSU &\scriptsize~~~2 &\scriptsize~~~3 &\scriptsize~~~16 &\scriptsize~~~0 &\scriptsize~~~0 &\scriptsize~~~3 &\scriptsize~~~0 &\textbf{\scriptsize~~~81} \\
		\cline{2-9}
	\end{tabular}
	\label{confusion2}
\end{table}

\section{Conclusion}
\label{conclusion}

RFI is a significant concern for modern radio astronomy, so the ability to identify the sources of RFI near radio telescope arrays is highly desirable. Once identified, RFI sources can be removed or replaced. Transient RFI as generated unintentionally by devices such as mechanical relays or fluorescent lights is especially difficult to identify, but once identified, potentially easy to mitigate. In this paper, we have demonstrated a novel approach to identifying the sources of transient RFI in the time domain. Our proposed approach is the first to make use of CNNs and bidirectional LSTMs to classify transient RFI by source. Applied to an existing dataset of 63130 individual transient signals recorded from 8 common sources of RFI, good classification accuracy is achieved. 

Our approach is well suited for future use with independent RFI monitoring stations at radio telescope arrays such as MeerKAT. In particular, since it only requires short recordings of individual transients, it is unaffected by limited recording time, a problem faced by some RFI recording systems \cite{wolfaardt2016}.

In future work, rather than identifying specific devices, we aim to classify sources by their physical components to permit more general source identification. For example, it may be possible to identify physical features such as mechanical contacts, brushes and inductive coils, among others. From their presence, the nature of an unknown device may be inferred. 

\section*{Acknowledgment}

The financial assistance of the South African SKA project (SKA SA) is hereby acknowledged. Opinions expressed and conclusions arrived at are those of the author and are not necessarily to be attributed to the SKA SA (www.ska.ac.za). 

\ifCLASSOPTIONcaptionsoff
  \newpage
\fi

\balance
%\bibliographystyle{IEEEtran}
%\bibliography{bibliography2}
%\section*{References}

\input{bibliography2.bbl}

\end{document}